\documentclass{llncs}

\usepackage{amsmath}
\usepackage{amsfonts}
\usepackage{amssymb}
\usepackage{graphicx}
\usepackage{multirow}
\usepackage[english]{babel}
\usepackage[square,numbers,sort&compress]{natbib}

\numberwithin{equation}{section}
\numberwithin{figure}{section}

\begin{document}

\title{Personalised Medicine: Taking a New Look at the Patient}
\titlerunning{Personalised Medicine}
\author{Marco Scutari}
\authorrunning{Marco Scutari}
\tocauthor{Marco Scutari}
\institute{Genetics Institute, University College London, United Kingdom.}

\maketitle

Personalised medicine strives to identify the right treatment for the right
patient at the right time, integrating different types of biological and 
environmental information. Such information come from a variety of sources:
omics data (genomic, proteomic, metabolomic, etc.), live molecular diagnostics,
and other established diagnostics routinely used by medical doctors \cite{ginsburg}. 
Integrating these different kinds of data, which are all high-dimensional,
presents significant challenges in knowledge representation and subsequent
reasoning \cite{streib,weston}. The ultimate goal of such a modelling effort
is to elucidate the flow of information that links genes, protein signalling
and other physiological responses to external stimuli such as environmental
conditions or the progress of a disease.

Omics data, which include single-nucleotide polymorphisms (SNPs), protein and
gene regulatory networks, are investigated using high-throughput platforms such
as the ones developed for the Human Genome project \cite{human1,human2}. Systems
biology studies the relationships among the elements in omics data as they 
change in the presence of genetic and environmental perturbations, extending
techniques that were previously used on a smaller scale \cite{sachs}. Such
knowledge can improve our ability to understand and predict the behaviour of
complex biological systems, but requires careful handling in integrating 
different sources. On its own, each type of data often contains too much noise
for single biological signals to be identifiable, much less their interplay. 
Pooling the information available across omics data (e.g. sequencing
and expression information about relevant genes, possibly under different 
treatment regimens) provides an option to increase statistical power and
produce reliable knowledge representation models \cite{ideker}. 

The role of live molecular diagnostics, and to some extent of traditional
diagnostics, is to complement omics data with longitudinal measures of the
patient's condition that are easier and cheaper to collect. Several examples
of the modelling and implementation techniques involved are covered in the
previous sections. Integrating such diagnostics is essential because genetic
information correlate only imperfectly with protein levels \cite{gygi},
which in turn are very noisy predictors of most pathologies. 

Applications of personalised medicine fall roughly in three groups. Firstly,
drug discovery and development can be made more efficient and effective
\cite{ginsburg}. On the one hand, omics data can provide feedback at early
stages of drug discovery by replacing the traditional trial-and-error approach
with a hypothesis-driven one based on a formal knowledge representation model.
On the other hand, omics data can also be used to improve clinical trial
design by guiding patients selection and stratification based on predicted
drug toxicity and non-responders profiles. This is likely to prove more
effective than defining populations in terms of race or ethnicity, since only
$5-10\%$ of the total human genetic variance occurs between different ethnic
groups \cite{cavallisforza} and boundaries between different populations
are often not clear.

Secondly, several aspects of the diagnostic process can be improved. For example,
the normal behaviour of a biological system can be better defined at the molecular
level than using non-specific clinical signs. As a result, pathologies can be
classified with greater precision based on a molecular taxonomy \cite{ginsburg};
previously unknown differences have been highlighted in breast cancer \cite{golub}
and leukaemia \cite{waring} in this way. Furthermore, genetic tests need to be
improved in their sensitivity and specificity; they are challenging to perform
reliably and interpret correctly, and they focus predominantly on rare diseases
\cite{hamburg}.

Thirdly, personalised medicine allows treatment for many diseases to be tailored
to each patient to an unprecedented degree. For example, adverse reactions to
specific compounds can be predicted with greater accuracy, and non-responders can
be identified without actually starting a therapy that may or may not be effective.

To investigate and implement personalised medicine in practice, many challenges
need to be overcome at the modelling level; some of them will be covered in the
following chapter. First and foremost, a working knowledge representation model
must be established to facilitate reasoning on high-dimensional, heterogeneous
data. Currently, probabilistic graphical models (Bayesian networks in particular)
seem to be a popular approach \cite{gene,mourad,cooper}. Their ability to
provide at the same time an intuitive understanding of the data to biologists
and medical doctors (through the graph structure) and a rigorous probabilistic 
framework to statisticians and computer scientists makes them an ideal tool for
this task.

Moreover, specific distributional assumptions are required to accurately describe
both omics and diagnostics data effectively. Gaussian and discrete Bayesian
networks from classic literature \cite{balding} present important limitations
in modelling omics data, as do more general models such as chain graphs. For
instance, assuming normality for gene expressions will almost certainly result
in a biased model, because expression levels are usually highly skewed.
Likewise, ignoring the ordering of the alleles in SNP data disregards
information which is known to be fundamental in quantitative genetics. Ideally,
probabilistic assumptions should also support the inclusion of available prior
information from different sources, as in Schadt et al. \cite{schadt}. 

A related issue is the computational complexity of both model estimation and
subsequent inference, which poses severe limits to the use of flexible 
distributional assumptions in Bayesian networks and to the scope of the 
questions these networks can answer. The use of prior information can speed up
model estimation by reducing the set of the models under consideration, even
though it may introduce bias as well if the phenomenon we are modelling is
not well understood. Another possible solution is to perform feature selection
as a pre-processing step, thus speeding up inference as well. In the context of
Bayesian networks, Markov blankets provide a natural way to do so while retaining
as much information as possible \cite{sahami}. However, given the complexity
of the data used in personalised medicine, the cost of feature selection is
often as high as that of model estimation.

In conclusion, while there are many open problems to address, an effective use
of knowledge representation is crucial in implementing reliable personalised
medicine protocols. Omics and other established diagnostics provide a wealth
of data, which calls for appropriate modelling spanning techniques from
statistics, computer science and quantitative biology.

\end{document}